\documentclass[a4paper,12pt]{article}
\pdfoutput=1
\usepackage{amsmath,amssymb,amsfonts,cite,color,epsf,epsfig,epstopdf,graphics,graphicx,mathtools,subcaption,physics,verbatim}
\usepackage{mathtools}
\def\thefootnote{\arabic{footnote}}
\definecolor{ultramarine}{rgb}{0.07, 0.04, 0.56}
\definecolor{cadmiumgreen}{rgb}{0.0, 0.42, 0.24}
\definecolor{indigo(dye)}{rgb}{0.0, 0.25, 0.42}
\usepackage[linktocpage=true,breaklinks]{hyperref}
\hypersetup{
	colorlinks=true,
	citecolor=ultramarine,
	linkcolor=cadmiumgreen,
	urlcolor=indigo(dye),
}

\numberwithin{equation}{section}

\usepackage{color}

\usepackage{amsmath, amssymb, graphics, epsfig, graphicx}
\usepackage{epsf}
\usepackage{epstopdf}
\usepackage {amssymb}

\graphicspath{{figs/}}

\newcommand{\nc}{\newcommand}
\nc{\ba}{\begin{eqnarray}}
\nc{\ea}{\end{eqnarray}}

\newcommand{\p}{{\partial}}
\newcommand{\Mpl}{M_{\rm Pl}}
\newcommand{\A}{A^T}
\newcommand{\m}{m_{\tiny A}}

\newcommand{\eq}[1]{\begin{equation}#1\end{equation}}

\newcommand{\fg}[1]{\begin{figure}[tbp]\centering #1 \end{figure}}

\nc{\e}{{\bf{e}}}

\begin{document}
	
	%%%%%%%%%%%%%%%%%%%%%%%%%%%%%%%%%%%%%%%%%%%%%%%%%%%%%
	%\begin{flushright} {\footnotesize IC/2007/001}  \end{flushright}
%	\vspace{5mm}
%	\vspace{0.5cm}
\begin{center}
		
\def\thefootnote{\fnsymbol{footnote}}

{ \bf  \large  Early Dark Energy and Dark Photon Dark Matter
	\\
	[.3cm]from Waterfall Symmetry Breaking}		
%{ \bf  \large  Waterfall Symmetry Breaking:\\
%			$H_0$ tension and/or dark matter origin}
\\[1cm]

{ Alireza Talebian$\footnote{talebian@ipm.ir}$}\\[0.5cm]

{\small \textit{ School of Astronomy, Institute for Research in Fundamental Sciences (IPM) \\ P.~O.~Box 19395-5531, Tehran, Iran }}\\
\end{center}

\vspace{.8cm}

\hrule \vspace{0.2cm}
%{\small  \noindent \textbf{Abstract} \\[0.3cm]
%\noindent

%%%%%%%%%%%%%%%%%%%%%%%%%%%%%%%%%%%%%%%%%%%%%%%%%%%%%

\begin{abstract}
	
	We investigate a cosmological model wherein a waterfall symmetry breaking occurs during the radiation-dominated era. The model comprises a complex waterfall field, an axion field, and the gauge field (dark photon) generated through a tachyonic instability due to the Chern-Simons interaction. Prior to symmetry breaking, the total energy density incorporates a vacuum energy 
	from the waterfall field, establishing a novel scenario for Early Dark Energy (EDE). Subsequent to the symmetry breaking, the dark photon dynamically acquires mass via the Higgs mechanism, potentially contributing to the  dark matter abundance. Hence, our model can simultaneously address the  $H_0$ tension and the origin of dark matter.  
	
\end{abstract}
\vspace{0.2cm} \hrule
\def\thefootnote{\arabic{footnote}}
\setcounter{footnote}{0}
\newpage
%\tableofcontents
%%%%%%%%%%%%%%%%%%%%%%%%%%%%%%%%%%%%%%%%%%%%%%%%%%%%%

\section{Introduction}
\label{sec:intro}

Current cosmological observations strongly support the flat Friedmann–Lemaître–Robertson–Walker (FLRW) spacetime model, which incorporates cold dark matter (CDM) and the cosmological constant ($\Lambda$), commonly referred to as the standard $\Lambda$CDM model. Despite constituting the predominant matter content in the universe, the exact identity of dark matter (DM) remains elusive. Various candidates have emerged, including weakly interacting massive particles, axion-like dark matter, vector dark matter, and primordial black holes~\cite{Jungman:1995df, Bertone:2004pz,Marsh:2015xka,Carr:2020xqk, Firouzjahi:2020whk,Hui:2016ltb,HosseiniMansoori:2020mxj,Hui:2021tkt,Hooshangi:2022lao,Nelson:2011sf,Firouzjahi:2021lov,Fabbrichesi:2020wbt,Talebian:2022cwk,Arias:2012az,Talebian:2022jkb,Dror:2018pdh,Gorji:2023cmz,HosseiniMansoori:2023mqh,Green:2020jor}.
%There are many dark matter candidates based on physics beyond the standard model %\cite{Vadim:2023QCDNonequil} Examples include axion-like dark matter, weakly interacting massive particles, vector dark matter, and primordial black holes.
While the fundamental nature of $\Lambda$ and DM remain an open question in cosmology, recent advancements in observational precision have brought to light several tensions in various observations. Among these, the most prominent is the Hubble tension—a significant discrepancy (at the $5\sigma$ level) between the current Hubble constant value derived from early-universe measurements, assuming the $\Lambda$CDM model, and the value directly measured from local observations~\cite{Verde:2019ivm,Riess:2019qba}. Specifically, the SH0ES team has obtained a value of $H_0 = 73.04 \pm 1.04 ~{\rm km/s/Mpc}$ (68\% confidence level) by using the HST observations of Cepheids and the Type-Ia supernova data~\cite{Riess:2021jrx}. In contrast, the current Hubble constant value inferred from Planck data, which relies on acoustic oscillations in the cosmic microwave background (CMB) power spectrum and assumes the $\Lambda$CDM model, is reported as $H_0 = 67.36 \pm 0.54~ {\rm km/s/Mpc}$ (68\% confidence level)~\cite{Planck:2018vyg}.

The origin of the Hubble tension remained as a big question for cosmologists. Although various potential sources of systematic effects have been considered~\cite{Davis:2019wet,Arendse:2019hev,Wojtak:2013gda,Odderskov:2017ivg,Wu:2017fpr}, the corresponding errors have not garnered a consensus among astronomers. Given the simplicity of local measurements of the Hubble constant, many proposed solutions involve introducing new physics to reconcile the value of $H_0$ inferred from CMB data. From this perspective, the tensions may be interpreted as indications of new physics beyond the framework of $\Lambda$CDM \cite{Kamionkowski:2022pkx, DiValentino:2021izs, DiValentino:2020zio,DiValentino:2018gcu,DiValentino:2016hlg,Reeves:2022aoi,Vagnozzi:2019ezj,Visinelli:2019qqu}.

One simple idea to increase  the value of the Hubble constant from the CMB data, is to slightly modify the evolution of the early universe 
%(the denominator in \eqref{H0}) 
without affecting the late universe. 
%(the numerator in \eqref{H0}). 
For an example, 
%One can postulate that 
the energy density before the last scattering is increased so that the expansion rate at the last scattering
%the denominator in \eqref{H0} 
is smaller than that in $\Lambda$CDM model\footnote{Another approach is to 
	%increase the numerator in \eqref{H0} by 
	decrease the energy density 
	%at the times 
	between now and at the time of decoupling. These approaches are known as late-time solutions which encounter other serious challenges (see Ref.~\cite{Kamionkowski:2022pkx} and references therein).}. In fact, it means we are suppressing the size of the sound horizon,
% $r_s$, 
	while keeping the peak heights in the CMB power spectrum and angular scales fixed through small shifts in other parameters. This is the basic idea behind early dark energy (EDE)~\cite{Poulin:2018cxd,Agrawal:2019lmo,Smith:2019ihp,Niedermann:2020dwg,Moshafi:2022mva}, a new ingredient which is added to the standard model.  
%The role of the EDE is to suppress the size of the sound horizon $r_s$, while keeping the peak heights and angular scales fixed through small shifts in other parameters. 
%More precisely, EDE is a new ingredient which is added to the standard model. 

The energy density of EDE behaves like a cosmological constant before some critical redshift $z_c$ and then it disappears quickly, like radiation or faster. %the critical redshift $z_c$. 
Therefore, the maximum contribution of EDE to the total energy density is at $z_c$ so it slightly enhances the expansion rate around that time. The EDE scenario can be realized by a single scalar field (e.g. axion) with transitions from a slow-roll phase (EDE phase) to an oscillating phase (decaying phase) through a second-order phase transition. 

A successful EDE cosmology can be simply parameterized by three parameters $\{
f_{\rm EDE}, z_c, n\}$ where $f_{\rm EDE}$ is the ratio of the EDE energy density $\rho_{\rm EDE}$ to the total energy density of universe  at $z_c$, $\rho(z_{c})$, and $n$ determines how quickly EDE component disappears after $z_c$, i.e. $\rho_{\rm EDE}(z<z_c) \propto a^{-n}$. Using the various dataset including the SH0ES, Planck, Pantheon, and BAO measurements, it is found out that cases with $f_{\rm EDE} \sim {\cal O}(5\%)$ and $z_c$ typically taken
around the matter-radiation (MR) equality works fine if the EDE component dilutes like radiation or faster $n\geqslant 4$. As an example, for oscillating-field and slow-roll models for EDE~\cite{Poulin:2018cxd}, the best-fit $\chi^2$ for the triplet $\{4.4\%,5345,4\}$ is reduced by -9 compared to $\Lambda$CDM. This means the EDE cosmology has a slight preference with respect to $\Lambda$CDM.

%As two examples, for oscillating-field and slow-roll models for EDE~\cite{Poulin:2018cxd}, the best-fit $\chi^2$ for triplets $(2.8\%,5345,4)$ and $(5\%,4965,4.5)$ are reduced by -9 and -14, respectively, compared to $\Lambda$CDM. This means the case $n=4.5$ has a slight preference with respect to $n=4$.

Although an axion field, with slow-roll and oscillation phases, is a candidate for the EDE scenario, some points must be considered. First of all, in order to agree with observations sensitive to the axion-like field perturbations, the initial field must be set close enough to the potential maximum, and the fluctuations must also be small enough. Second, without fine-tuning of the axion potential the EDE energy density does not dilute fast enough after the transition to be in agreement with data. There are new proposals for EDE scenario which it is claimed that they can address the fine-tuning issues of the old EDE model \cite{Niedermann:2019olb,Niedermann:2020dwg,Niedermann:2020qbw,Niedermann:2021ijp,Niedermann:2021vgd, Nakagawa:2022knn}

In this paper, we propose a new scenario for EDE which is based on waterfall phase transition~\cite{Linde:1993cn,Copeland:1994vg}. Historically, this mechanism was proposed for ending the inflationary era where instead of oscillation of inflaton or a first order phase transition, inflation ends due to a very rapid rolling (‘waterfall’) of a complex scalar field. %triggered by inflaton. 
This mechanism is a hybrid of chaotic inflation and the theory of spontaneous symmetry breaking and then differs both from the slow-rollover and the first-order inflation. 

The waterfall field, which is responsible for the symmetry breaking, can be charged under $U(1)$ gauge field~\cite{Emami:2013bk,Abolhasani:2013bpa}. Following that, we referred to the gauge field as ``\textit{dark photon}''. Here we employ this mechanism during the radiation-dominated (RD) era for terminating EDE phase. As expected, there is no need for 
%elaborate 
a contrived elaborate potential or fine-tuning with respect to the initial condition. More interestingly, after the waterfall becomes tachyonic at the end of EDE, rolling rapidly towards its global minimum, the dark photon acquires mass via the Higgs mechanism which can contribute to the observed dark matter abundance. Therefore, our model can  address simultaneously both problems of $H_0$ tension and the origin of DM.

%{\color{red} 
	The rest of the paper is organized as follows.%} 
In Section \ref{sec:model} we review the setup of waterfall phase transition including a complex waterfall field, a real scalar field, and a $U(1)$ gauge field. In Section \ref{sec:waterfall_RD} we implement our model when the transition happened during RD era and derive evolution of the real scalar field (axion) and calculate the relic abundance of the gauge field. In Sections \ref{sec:EDE}, we explore the parameter spaces of the model to alleviate $H_0$ tension and to address the origin DM.
% via the generated dark photon, respectively. 
We end with discussion and conclusions in Section \ref{sec:summary}.

%%%%%%%%%%%%%%%%%%%%%%%%%%%%%%%%%%%%%%%%%%
\section{The Model}\label{sec:model}

In this section we review our model which is based on waterfall phase transition. It contains a complex scalar field $\psi$ (the waterfall field), a real scalar field $\phi$ (the axion field), and the dark photon $A_\mu$. This picture is similar to the setup employed in \cite{Salehian:2020asa} 
which studied the generating of dark photon dark matter from inflationary perturbations. The action is given by \cite{Emami:2011yi,Abolhasani:2013bpa}
\begin{eqnarray}
\label{action} 
{\cal S} = \int
\dd^4 x \,  \sqrt{-g} \bigg[ \frac{\Mpl^2}{2} \mathcal{R} - \frac{1}{2}(\p\phi)^2
- \frac{1}{2} |D\psi|^2 - V(\phi,|\psi|) 
- \frac{1}{4}  F_{\mu \nu} F^{\mu \nu} - \frac{\alpha \phi}{4f_{\rm a}}  F_{\mu \nu} {\tilde F^{\mu \nu}} \bigg] \,,
\end{eqnarray}
where $\Mpl = 1/ \sqrt{8 \pi G}$ is the reduced Planck mass with $G$ being the Newton constant and $\mathcal{R}$ is the Ricci scalar associated with the spacetime metric $g_{\mu\nu}$. The complex waterfall  field $\psi$ is charged under the $U(1)$ gauge field $A_\mu$ with the covariant derivative  given by 
\ba
D_\mu
\psi = \partial_\mu  \psi + i \e \,  \psi  \, A_\mu \,,
\ea
where $\e$ is the dimensionless gauge coupling between the dark photon and the waterfall field. The components of
the field strength tensor associated to the gauge field is given by $F_{\mu \nu}=\partial_\mu A_\nu - \partial_\nu A_\mu$ and $\tilde{F}^{\alpha\beta}=\eta^{\alpha\beta\mu\nu}F_{\mu\nu}/\left(2\sqrt{-g}\right)$ is its dual with $\eta^{0123}=1$. Finally, the axion is coupled to the dark photons through the Chern-Simons interactions. The coupling is determined by $\alpha/f_{\rm a}$ in which $f_{\rm a}$ is the axion decay constant while $\alpha$ is a dimensionless constant.

We assume the scalar potentials depend on the modulus of the scalar fields. This can be supported by imposing the axial symmetry. One can decompose the  waterfall field into the radial and angular components, $\psi \equiv |\psi| e^{i \theta}$, and exploit the gauge symmetry to set $\theta=0$, i.e. going to the unitary gauge.  In this gauge, $\psi$ is practically real, and we use the notation $|\psi|\equiv \psi$ afterwards. Our model is based on 
%We consider 
a Higgs-like symmetry breaking potential as in hybrid inflation \cite{Linde:1993cn,Copeland:1994vg}:
\begin{equation}
\label{pot}
V(\phi,\psi) = \frac{\lambda}{4} \Big( \psi^2 - \frac{M^2}{\lambda}\Big)^2 
+ \frac{1}{2} m^2 \phi^2 + \frac{1}{2} {g}^2 \phi^2\psi^2 \, ,
\end{equation}
where $\lambda$ and $g$ are two dimensionless couplings and $M$ is a mass scales, controlling the mass of the waterfall field. The symmetry breaking triggers the Higgs mechanism, inducing mass for $A_\mu$

The overall picture of the dynamics is as follows. Let us define the critical value of the axion field $\phi_c$ via $\phi(z_c)=\phi_c \equiv M/g$. The dynamics has two stages. At early stages $z>z_c$, the axion $\phi >\phi_c$ is rolling slowly towards $\phi_c$ while the waterfall field is very heavy, rapidly rolling to its local minimum $\psi=0$. 
We assume the potential \eqref{pot} is mainly dominated by its vacuum during this stage so it provides a period of acceleration expansion, e.g. an inflationary Universe or a period of EDE. This assumption requires that 
\ba
\label{rho_axion}
\frac{M^4}{4 \lambda} \gg \frac{1}{2} m^2 \phi^2 \quad  \longrightarrow\quad    M^2 \gg \frac{\lambda}{g^2} m^2  . 
\ea
Hence, the total energy density includes a constant contribution from the waterfall field, serving as a form of EDE:
\begin{align}
\label{rho_psi}
\rho_{\rm EDE} = \dfrac{M^4}{4\lambda} \equiv f_{\rm EDE} ~ \rho(z_{\rm c}) \,,  
\end{align}
where $f_{\rm EDE}<1$ and $\rho_{\rm c}$ %\equiv \rho_{\rm tot} (z_c)$ 
denotes the total energy density at the end of EDE phase. Furthermore, given our intention to position the EDE phase (which coincides with the end of symmetry breaking) close to the matter-radiation equality epoch, $z_c \sim z_{\rm eq}$, it is necessary to consider the contribution of matter as well:
\begin{align}
\label{rho-e}
\rho(z_{\rm c})  = 3\Mpl^2H_c^2 = \dfrac{\pi^2}{30}g_{*c} T_c^4\left(
1+\dfrac{a_c}{a_{\rm eq}}
\right) \,,
\end{align}
in which $H_c$, $T_c$, and $g_{*,c}\simeq 3.363$ are the Hubble expansion rate, the temperature of radiation fluid, and the effective relativistic degrees of freedom for energy density at $z_c$.

When the axion field reaches $\phi = \phi_c$, the waterfall field becomes tachyonic, initiating a rapid roll towards its global minimum at $\psi_{\rm min} = M/\sqrt{\lambda}$. Under the assumption of a significantly heavy waterfall mass, the transition to the global minimum and symmetry breaking occurs swiftly. 

The waterfall mechanism plays two crucial roles. First, it terminates the period of EDE,
%the symmetry breaking, 
similar to the termination of inflation in conventional hybrid inflation. Second, it induces mass to the dark photon. Similar to the standard Higgs mechanism, the gauge field obtains mass through the gauge coupling  $\e$ in the following manner:
\ba
\label{mA}
\m^2 =  \frac{\e^2 M^2}{\lambda}\,,\qquad\qquad(\text{at the end of symmetry breaking})\,.
\ea
However, it is important to note that the dark photon remains massless during the initial transition and only becomes massive in the final stage, i.e., after symmetry breaking. The massive dark photon drags the vacuum energy from the potential, subsequently assuming the roles of dark matter, or a significant part of it. Given that the energy density of dark matter is critically tied to the induced mass of the dark photon, we define the parameter $R$ as follows:
\eq{
	\label{R}
	R\equiv\frac{\m^2}{H_c^2} 
	%= \textcolor{red}{12 f_{\rm EDE}} \e^2  \big( \frac{M_P}{M} \big)^2 
	\, .
}
Parameter $R$ effectively quantifies the mass of the dark photon relative to the Hubble expansion rate at the moment of dark photon generation. By combining equations \eqref{mA}, \eqref{R}, and \eqref{rho_psi}, we arrive at the following expression:
\begin{align}
\label{R_2}
R = \dfrac{\e^2}{\sqrt{\lambda}}\sqrt{12f_{\rm EDE}}\dfrac{\Mpl}{H_c} \,,
\end{align}
which can be very large when considering typical values for $\e, \lambda$ and $f_{\rm EDE}$ during RD era.

In addition, after symmetry breaking, a large mass is induced for the axion via the coupling $g^2 \psi^2$. We require that this induced mass is significantly greater than $H_c$, ensuring that the axion rapidly oscillates around its minimum at $\phi=0$ to prevent the generation of a second stage of dark energy at a later time. 

%%%%%%%%%%%%%%%%%%%%%%%%%%%%%%%%%%%%%%%%%
\iffalse
\fg{
\centering
\includegraphics[width=0.55\textwidth]{hybrid1}
\caption{\scriptsize A schematic view of symmetry breaking to terminate EDE (adapted from \cite{Salehian:2020asa}).   The red line is the chaotic potential for the axion $\phi$ while the blue curve is the potential of $\psi$.  Note that we always have $\psi\geq0$.}
\label{fig:hybrid}
}
\fi
%%%%%%%%%%%%%%%%%%%%%%%%%%%%%%%%%%%%%%%%

One interesting prediction of our setup is the production of cosmic strings at the end of symmetry breaking. This is a general consequence of $U(1)$ symmetry breaking in 
early universe \cite{Kibble:1976sj, Vilenkin:2000jqa,Felder:2000hj,Felder:2001kt}. Unlike monopoles and domain walls, cosmic strings can form at various points in cosmic history, resulting in intriguing observational effects, such as lensing, the generation of CMB anisotropies, or the production of stochastic gravitational waves \cite{Vilenkin:2000jqa}. The tension of the strings, denoted by $\mu$, is determined by the scale of symmetry breaking \cite{Vilenkin:2000jqa}:
\ba
\label{string-mu}
\mu \sim \frac{M^2}{ \lambda} \, .
\ea
The bounds from CMB anisotropies require that  $G \mu \lesssim 10^{-7}$ \cite{Planck:2013mgr}. In our analysis, we will demonstrate that the cosmic strings produced in our setup do not approach this upper bound, and as a result, their effects can be safely neglected.

\subsection{Dark photons Production }\label{sec-DP}

As previously discussed, the gauge fields are generated through a tachyonic instability as the axion field undergoes a rolling phase and dynamically acquires mass via symmetry breaking at the end of the phase transition. We will now provide a brief overview of this process.

We work with a spatially flat FLRW metric, 
\begin{align}\label{FLRW}
	{\rm d}s^2 = a^2(\tau) \big(-{\rm d}\tau^2+\delta_{ij}~{\rm d}x^i{\rm d}x^j \big) \, ,
\end{align}
where the scale factor is denoted as $a(\tau)$, and $\tau$ represents the conformal time, related to cosmic time through the standard relation $\dd t = a(\tau) \dd\tau$.

Given the action \eqref{action}, the energy-momentum tensor and the equation of motion for the gauge field can be represented as
\begin{align}
\label{EMT-EM}
&T_{\sigma\rho}= F_{\sigma\eta}F_\rho{}^\eta-\frac{1}{4}g_{\sigma\rho}F_{\eta\delta}F^{\eta\delta}
+\m^2\Big(A_\sigma A_\rho-\frac{1}{2}g_{\sigma\rho}A_\eta A^\eta\Big)\,,
\\
\label{Maxwell}
&\nabla^\rho \big(  F_{\rho\sigma} + \frac{\alpha}{f_{\rm a}} \phi {\tilde F}_{\rho\sigma} \big) - \m^2 A_\sigma = 0 \,.
\end{align}
The gauge field lacks a classical background value, $\ev{A_\sigma}=0$ in our setup. Consequently, the temporal component of Eq.~\eqref{Maxwell} results in
\eq{\label{Eq.long}
	\big(\nabla^2-\m^2a^2\big)A_0=(\p_iA_i)'\,,
}
in which a prime denotes the derivative with respect to the conformal time. As usual,  $A_0$ is not dynamical, it  is a constraint and its solution can be imposed at the level of the action. Taking into account the background $O(3)$ symmetry, we can decompose the spatial components of the gauge field as $A_i=\partial_i\chi+A^T_i$, where $\chi$ represents the longitudinal part and $A^T_i$ corresponds to the transverse part of the vector field ($\partial_i A^T_i=0$). Substituting this decomposition into Eq.~\eqref{Eq.long}, we derive a solution for $A_0$ in favour of the longitudinal mode $\chi$. As demonstrated in  \cite{Salehian:2020asa}, the energy density contribution of the longitudinal mode is subleading when compared to the transverse mode. Therefore, in our subsequent analysis, we focus solely on the two transverse modes, leading to the following energy density:
\eq{
	\label{rhoT}
	\rho_T=\frac{1}{2a^4}\left[(\A_i{}')^2+(\p_i\A_j)^2+{\m^2a^2}(\A_i)^2\right]\,,
}
Therefore, the effective action is expressed as
\ba
\label{ST}
{\cal S}_T=\frac{1}{2}\int\dd[3]{x}\dd{\tau}\Big[  (\A_i{}')^2
-  (\p_i\A_j)^2-\m^2a^2(\A_i)^2 + \frac{\alpha \phi}{f_{\rm a}} \epsilon_{ijk} \A_i \p_j \A_k \Big] \, . 
\ea
To proceed further, we decompose the gauge field perturbations in terms of the creation and annihilation operators $a_{\vb{k}}$ and $a^\dagger_{\vb{k}}$ as:
\begin{equation}
\label{V-mode}
\A_i(\tau,\vb{x})=\sum_\lambda\int\frac{\dd[3]{k}}{(2\pi)^{3/2}} \varepsilon^\lambda_i({\bf k})
\big[v_{k,\lambda}(\tau)a_{\vb{k},{\lambda}}
+v_{k,\lambda}(\tau)^*a^\dagger_{-\vb{k},{\lambda}} \big] e^{i\vb{k}.\vb{x}}\,, \hspace{1cm} 
[a_{\vb{k},{\lambda}}, a^\dagger_{\vb{k}',{\lambda'}}] = \delta_{\lambda\lambda'}\delta(\vb{k}-\vb{k}')\,,
\nonumber
\end{equation}
where the time dependence of the gauge field is described by the mode functions $v_{k,\lambda}(\tau)$. In the above relation, $\varepsilon^\lambda_i({\bf k})$ are the polarization vectors for $\lambda =\pm$ which satisfy $\varepsilon_i^{\lambda}({\bf k}){}^*=\varepsilon_i^{-\lambda}({\bf k})=\varepsilon_i^{\lambda}(-{\bf k})$, $k_i\varepsilon_i^\lambda({\bf k})=0$, and also the identity $\epsilon_{ij\ell}k_j\varepsilon_\ell^\lambda({\bf k})=-i\lambda{k}\varepsilon_i^\lambda({\bf k})$. (see appendix A of Ref. \cite{Salehian:2020dsf} for more details).

Before symmetry breaking, the gauge field is massless, so the equation for the mode function is as follows
\ba\label{MF-Eq}
v_{k,\lambda}'' + \Big( k^2 -   \frac{\lambda\alpha }{f_{\rm a}}\phi' k \Big) v_{k,\lambda} =0  \,, 
\qquad\text{(before symmetry breaking: $z>z_c$) \,.}
\ea
Note that during the rolling of axion ($\phi' \neq 0$), the mode functions for different polarizations evolve differently. For $\phi'<0$, the negative-helicity modes with $k<\alpha |\phi'|/f_a$ experience a tachyonic instability and grow non-perturbatively while the positive-helicity modes damp exponentially.

Crucial to our discussion is that  at the end of symmetry breaking, the dark photon becomes massive. Denoting the mode function of the massive vector field as $u_{k,\lambda}$, its equation of motion reads
\ba\label{MF-Eq2}
u_{k,\lambda}'' + \Big( k^2 -   \frac{\lambda\alpha \phi'}{f_{\rm a}} k
+ {\m^2a^2} \Big) u_{k,\lambda} =0\,,
\qquad\text{(after symmetry breaking: $z<z_c$)}\,.
\ea
We are interested in calculating the energy density of the dark photons after they acquire mass, which, from Eq.~\eqref{rhoT}, is given by
\eq{
	%\label{rhoT4}
	\rho^{(\tiny A)}(a)=\frac{1}{2a^4} \sum_{\lambda} \int\frac{\dd[3]{k}}{(2\pi)^3}
	\Big[ \big|{{u_{k,\lambda}}'}\big|^2+\big(k^2+{\m^2a^2}\big)\abs{u_{k,\lambda}}^2\Big]\,. 
} 
By imposing the matching condition at the surface of the phase transition between the two mode functions, $u_{k,\lambda}(\tau_c)=v_{k,\lambda}(\tau_c)$ and $u'_{k,\lambda}(\tau_c)=v'_{k,\lambda}(\tau_c)$, we finally arrive at
\eq{
	\label{rhoT5}
	\rho^{(\tiny A)}(z_{\rm c}) = \frac{1}{2a_c^4} \sum_{\lambda} \int\frac{\dd[3]{k}}{(2\pi)^3}
	\left[ \big| {{v_{k,\lambda}}'}\big|^2+\big(k^2+{\m^2a^2}\big)\abs{v_{k,\lambda}}^2\right] \bigg|_{\tau_c}\,,
} 
for the energy density of the gauge field at the end of symmetry breaking. Equation \eqref{rhoT5} effectively represents the accumulated energy of dark photons, encompassing induced mass effects, after the symmetry breaking.

Drawing inspiration from \cite{Salehian:2020dsf}, we can represent the energy density of the generated dark photons using the following parameterization 
\begin{equation}
	\label{rho-param}
\rho^{(\tiny A)}(z_{\rm c}) \equiv {\cal C}~ \rho_{c}(z_{\rm c})= 3 {\cal C}\Mpl^2H_c^2
\,,
\end{equation}
where the parameter ${\cal C}$ is the fractional energy density of dark photon and $\rho_c$ is the total energy density at $z_c$. This energy dilutes as $a^{-4}$ or $a^{-3}$, depending on whether the dark photon is relativistic or non-relativistic, respectively. 

In section \ref{sec: relic}, we will derive an expression for ${\cal C}$ in term of the parameters of the model.

%%%%%%%%%%%%%%%%%%%%%%%%%%%%%%%%%%%%%%%%%%
\subsection{Constraints on the model}

There are two sources of backreaction from dark photons that must be negligible in order for model to be consistent. The first is the backreaction on the geometry, and the second is the backreaction on the dynamics of the axion. The former condition is that the energy density of dark photons must be small compared to the total energy density before symmetry breaking, i.e. $\rho_c^{(\tiny A)}\ll 3\Mpl^2H_c^2$. This condition is easily satisfied as
\eq{\label{BR-1}
	{\cal C} \ll 1 \,.
}
Second, the contribution of the dark photons to the axion field equation must be small. The equation of motion for the axion field is as follows:
\eq{
	\label{KG}
	\ddot{\phi}+3H\dot{\phi}+\dv{V}{\phi}=\frac{\alpha}{f_a} E_i B_i \,,
}
where $E_i = - a^{-2} (A'_i - \p_i A_0)$ and $B_i = a^{-2} \epsilon_{ijk} \p_j A_k$ are the corresponding electric and magnetic fields of the dark gauge field. The backreaction condition from the axion field equation requires
\eq{\label{BR-2}
	S_\phi \ll 1 \,;
	\hspace{2cm}
	S_\phi \equiv \abs{\frac{\alpha}{f_a}\dfrac{ E_i B_i }{3H\dot{\phi}}}
	\,.
}
For the later purposes, it is convenient to express $S_\phi$ for tachyonic modes associated with the negative-helicity in terms of the mode functions at the end of symmetry breaking, as follows
\begin{align}
\label{SSS}
S_\phi(\tau_{\rm c}) &= \frac{1}{3a_{\rm c}^4H_{\rm c}\abs{\dot{\phi}_{\rm c}}}\dfrac{\alpha}{f_a}\int\frac{ k^3\dd{k}}{2\pi^2} \Re \Big[ v_{k,-}^* {v'_{k,-}} \Big]_{\tau_c} \,.
%\ll 1
\end{align}
The conditions \eqref{BR-1} and \eqref{BR-2} must be satisfied for the model to be consistent. It is important to note that both of these conditions should always hold during the radiation era. However, given that the gauge field energy density is increasing, these two conditions are most stringent at the end of the EDE phase. Therefore, we will evaluate them at the redshift $z_c$.
%%%%%%%%%%%%%%%%%%%%%%%%%%%%%%%%%%%%%%%%%%%%%%%%%%%%

\section{Waterfall Phase Transition during RD era}
\label{sec:waterfall_RD}

In this section, we adopt the approach outlined in \cite{Salehian:2020asa} for dark photon production through the waterfall mechanism. However, our analysis differs in that we consider the radiation-dominated (RD) era, where the relationship between the scale factor and time is given by
\begin{align}
\dfrac{a}{a_i}=\dfrac{\tau}{\tau_i}=\Big(
\dfrac{t}{t_i}
\Big)^{1/2} \,,
%\\
%a(t) = a_i \Big(
%\dfrac{t}{t_i}
%\Big)^{1/2} \,,
%\hspace{.5cm}
%\tau(t) = \tau_i \Big(
%\dfrac{t}{t_i}
%\Big)^{1/2} \,,
%\hspace{0.5cm}
%\tau_i a_i  = 2t_i\,,
%\hspace{0.5cm}
%H(t) = \dfrac{1}{2t} \,,
%\hspace{.5cm}
%a(t) ~ \tau(t) ~ H(t) =1
\end{align}
where $\tau_i a_i  = 2t_i$. Here we have ignored the possible contribution of matter to the scale factor close to the matter-radiation equality. In the following sub-sections, we will study the evolution of the axion field during RD era and provide an estimate for the relic abundance of the generated dark photons.

\subsection{Evolution of axion field}
\label{sec-axion}
Prior to the symmetry breaking, the axion field has a tiny mass $m \ll H$, allowing it to roll down slowly before the phase transition. Neglecting the backreaction of the gauge field on the dynamics of the axion, as indicated in \eqref{BR-2}, and we verify it below, the solution of \eqref{KG} is given by
\begin{align}
\label{phi_t}
\phi(t) = t^{-1/4} \Big(
c_1 J_{1/4}( m t) + c_2 Y_{1/4}( m t)
\Big) \,,
\end{align}
where $J$ and $Y$ represent the Bessel functions of the first and second kinds, respectively. The coefficients $c_1$ and $c_2$ can be determined through initial or boundary conditions. Assuming a regular solution as $t \rightarrow 0$, we find that $c_2=0$. Introducing the dimensionless parameter $x \equiv m t$, we can approximate the solution in \eqref{phi_t} as follows
\begin{align}
\label{phi_x_approxi}
\phi(x) \simeq \dfrac{\phi_0}{5} (5-x^2) \,,
\end{align}
where $\phi_0=\phi(x=0)$ is chosen for the initial value of axion field. To prevent the oscillatory behavior of $\phi$ until just before the symmetry breaking, it is necessary that $x_c=mt_c <1$, where $t_c$ is determined by $\phi(t_c)=\phi_{\rm c}$. Using \eqref{phi_x_approxi}, we obtain the following relation:
\begin{align}
\label{phi0phic}
\phi_{\rm c} < \phi_0 < 1.25 \phi_{\rm c} \,,
\end{align}
where we have used 
\begin{align}
\label{phi_cc}
x_c &=m t_c \simeq \sqrt{5\left(
	1-\dfrac{\phi_c}{\phi_0}
	\right)} \,.
\end{align}
It's worth estimating the e-folding number, denoted as $N_c$, which corresponds to the e-folding time interval from $t_i$ (the onset of $\phi$ rolling) to $t_c$ (the trigger time), i.e. $N_c \equiv \ln(a_c/a_i)$. Utilizing the relation $x_c = mt_i \left(a_c/a_i\right)^2$, we obtain
\begin{align}
\label{Ne}
e^{2N_c} \simeq 2\dfrac{H_i}{m}\sqrt{5\left(
	1-\dfrac{\phi_c}{\phi_0}
	\right)} \,,
\end{align}
where $H_i$ is the Hubble expansion rate at $t_i$. In Fig.~\ref{fig:Ne} we have shown a density plot of $N_c$ in term of $H_i/m$ within the allowed regime \eqref{phi0phic}. Furthermore, there is an upper bound on $N_c$ as follows.
%we aim to apply the waterfall mechanism for the EDE scenario. 
Assuming the EDE phase began after BBN and concluded before matter-radiation (MR) equilibrium, we obtain
\begin{align}
N_{\rm eq}-N_{\rm BBN} %= \ln(\dfrac{a_{\rm eq}}{a_{\rm BBN}}) 
\simeq \ln(\dfrac{T_{\rm BBN}}{T_{\rm eq}}) %\simeq \ln(\dfrac{\rm MeV}{\rm eV}) \simeq 6 \ln(10) 
\simeq 15 \,.
\end{align}
Hence, there exists an upper bound for the number of $e$-foldings during the rolling phase of $\phi$, which is given by $N_c < 15$. We will demonstrate that, for our proposed model, a value of $N_c \sim \mathcal{O}(1)$ is appropriate. The precise value depends on the couplings $\lambda$, $g$, and $\alpha$.

%%%%%%%%%%%%%%%%%%%%%%%%%%%%%%%%%%%%%%%%%
\fg{
	\centering
	\includegraphics[width=0.55\textwidth]{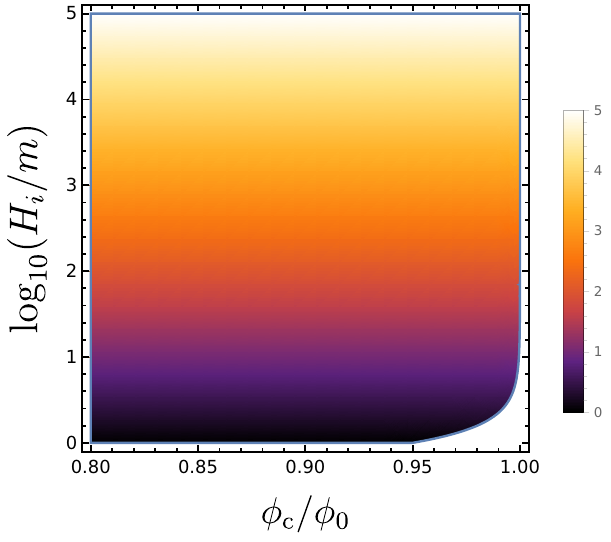}
	\caption{%\scriptsize 
		Density plot for $N_c$ using \eqref{Ne}. The forbidden regime located at bottom right is due to $N_c<0$.}
	\label{fig:Ne}
}
%%%%%%%%%%%%%%%%%%%%%%%%%%%%%%%%%%%%%%%%

%%%%%%%%%%%%%%%%%%%%%%%%%%%%%%%%%%%%%%%%%%
\subsection{Relic Abundance of Dark Photons}
\label{sec: relic}

In the presence of the Chern-Simon coupling term in action \eqref{action}, the dark photon quanta exhibit tachyonic instability, which is sourced by the rolling axion field $\phi$. More precisely, before the trigger time, the axion field $\phi$ rolls slowly and amplifies one polarization of the dark photon as described by \eqref{MF-Eq}. Depending on the value of the parameter $R$, the dark photons evolve either like radiation or matter afterward. The parameter space of our setup enables us to generate a very massive dark photon at the end of symmetry breaking. 

We will solve the mode function \eqref{MF-Eq} for the dark photon perturbations and calculate the parameter ${\cal C}$, as defined in \eqref{rho-param}. This will allow us to determine $\Omega_{A}$, representing the relic dark photon energy density as a component of or the entirety of the dark matter energy density.

By introducing the variable $y=k\tau$ and considering the negative-helicity case ($\lambda = -1$), the mode function given in Eq.~\eqref{MF-Eq} can be expressed as follows:
\begin{align}
\label{v-y}
\dfrac{\dd^2 v_k}{\dd y^2}+\bigg[
1-\Big(
\dfrac{y}{y_c}
\Big)^3
\bigg]v_k = 0 \,,
\end{align}
in which $v_k=v_{k,-}$ and
\begin{align}
\label{yc}
y_c (k) \equiv \Bigg(
\dfrac{4}{5}\dfrac{\alpha}{f_a}\phi_0\dfrac{x_i^2}{k^4 \tau_i^4}
\Bigg)^{-1/3}
\,.
\end{align}
To solve the equation \eqref{v-y}, we employ the matching condition technique. For $y \ll y_c$, the solution is given by
\begin{align}
	\label{v_1}
	%\dfrac{\dd^2 v^{(1)}_{k}}{\dd y^2} +  v^{(1)}_{k} = 0
	%\hspace{1cm} \longrightarrow \hspace{1cm}
	v^{(1)}_{k}(y) = \dfrac{e^{-i y}}{\sqrt{2 k}} 
\end{align}
where we have assume Bunch–Davies vacuum as the initial condition. For $y \gg y_c$ the solution takes the following form
\begin{align}
\label{v_2}
%\dfrac{\dd^2 v^{(2)}_{k}}{\dd y^2} - \big(
%\dfrac{y}{y_c}
%\big)^3 v^{(2)}_{k} = 0
%\hspace{0.5cm} \longrightarrow
%\hspace{0.5cm}  
	v^{(2)}_{k}(y) =  \sqrt{y}~\Bigg(
	d_1~ I_{1/5}\bigg(\frac{2}{5}\dfrac{y^{5/2}}{y_c^{3/2}}\bigg)+d_2~K_{1/5}\bigg(\frac{2}{5}\dfrac{y^{5/2}}{y_c^{3/2}}\bigg)\Bigg) \,,
\end{align}
where $I$ and $K$ are the modified Bessel functions of the first and second kinds, respectively. The constant coefficients $d_1$ and $d_2$ can be determined using the matching condition. We require both $v_{k}^{(1,2)}$ and $\partial_y v^{(1,2)}_{k}$ to be continuous at $y_c$. Finally, we arrive at
\begin{align}
\label{d1}
	d_1 &= \dfrac{\sqrt{2y_c}}{5\sqrt{k}}e^{-iy_c}\Bigg(
	K_{4/5}\bigg(\frac{2}{5}y_c\bigg)-i~K_{1/5}\bigg(\frac{2}{5}y_c\bigg)\Bigg)
	\,,
	\\
	\label{d2}
	d_2 &= \dfrac{\sqrt{2y_c}}{5\sqrt{k}}e^{-iy_c}\Bigg(
	I_{-4/5}\bigg(\frac{2}{5}y_c\bigg)+i~I_{1/5}\bigg(\frac{2}{5}y_c\bigg)\Bigg)
	\,.
\end{align}
The definition of $y_c$ in \eqref{yc} can be recast into the following form
\begin{align}
\label{y_c}
y_c^3 &= y_{\rm max}^{-1} ~y^4\,,
\\
\label{ymax}
y_{\rm max}(t) &\equiv \dfrac{1}{5}\dfrac{\alpha}{f_a}\phi_0 \Big(
\dfrac{m}{H(t)}
\Big)^{2} \,.
\end{align}
Utilizing the parameterization mentioned above, the tachyonic instability condition occurs for $y<y_{\rm max}$. Consequently, based on \eqref{rhoT5} and \eqref{SSS}, we can express the energy density of the dark photons and their influence on the evolution of $\phi$ as follows
\begin{align}
\rho^{(\tiny A)}(z_{\rm c}) &= \dfrac{H_c^4}{4\pi^2}\bigg(
I_1(y_{\rm max})+R~ I_2(y_{\rm max})
\bigg) \,,
\label{rho}
\\
S_\phi &= \dfrac{1}{6\pi^2}\bigg(
\dfrac{\alpha}{f_a}\Mpl
\bigg)^2 \left(
\dfrac{H_c}{\Mpl}
\right)^2~I_3(y_{\rm max})\,,
\label{S}
\end{align}
where $R$ was defined in \eqref{R} and
\begin{align}
\label{I1}
I_1(y_{\rm max}) &\equiv \int_{0}^{y_{\rm max}} y^4 \dd y \Bigg( \abs{\dfrac{1}{\sqrt{\tau}}\dfrac{\dd v^{(2)}_{k}}{\dd y}}^2 + \abs{\dfrac{v^{(2)}_k}{\sqrt{\tau}}}^2
\Bigg)_{\tau_c}\,,
\\
\label{I2}
I_2(y_{\rm max}) &\equiv \int_{0}^{y_{\rm max}} y^2 \dd y~\abs{\dfrac{v^{(2)}_k}{\sqrt{\tau}}}^2_{\tau_c}\,,
\\
\label{I3}
I_3(y_{\rm max}) &\equiv \dfrac{1}{y_{\rm max}}\int_{0}^{y_{\rm max}} y^4 \dd y~\Re \Bigg[ \dfrac{v_{k}^{(2)*}}{\sqrt{\tau}}\dfrac{1}{\sqrt{\tau}}\dfrac{\dd v^{(2)}_{k}}{\dd y} \Bigg]_{\tau_c}\,.
\end{align}
We have plotted $I_1, I_2$, and $I_3$ in terms of $y_{\rm max}$ in the left panel of Fig.~\ref{fig:I1I2_and_I1OvereI2}. 
%%%%%%%%%%%%%%%%%%%%%%%%%%%%%%%%%%%%%%%%%%%%%%%%%%%%%%%%%%%%%%%%%%%%%%%%%%%%%%%%%%%%%%%%%%%%%%%%%%%%%%%%%%%%%%%%%%%%%%%%%%%%%%%%%%%%%%%%%%%%%%%%%%%%%%%%%%%%%%%%%%%%
%\textcolor{red}{contribution of superhorizon modes of gauge field to DM??? For computing \eqref{rhoT5}, contribution of which modes must be considered?}
%%%%%%%%%%%%%%%%%%%%%%%%%%%%%%%%%%%%%%%%%%%%%%%%%%%%%%%%%%%%%%%%%%%%%%%%%%%%%%%%%%%%%%%%%%%%%%%%%%%%%%%%%%%%%%%%%%%%%%%%%%%%%%%%%%%%%%%%%%%%%%%%%%%%%%%%%%%%%%%%%%%%%%%%%%%%%%%%%%%%%%%%%%%%%%%%%%%%%%%%%%%%%
\fg{
	\centering
	\includegraphics[width=0.48\textwidth]{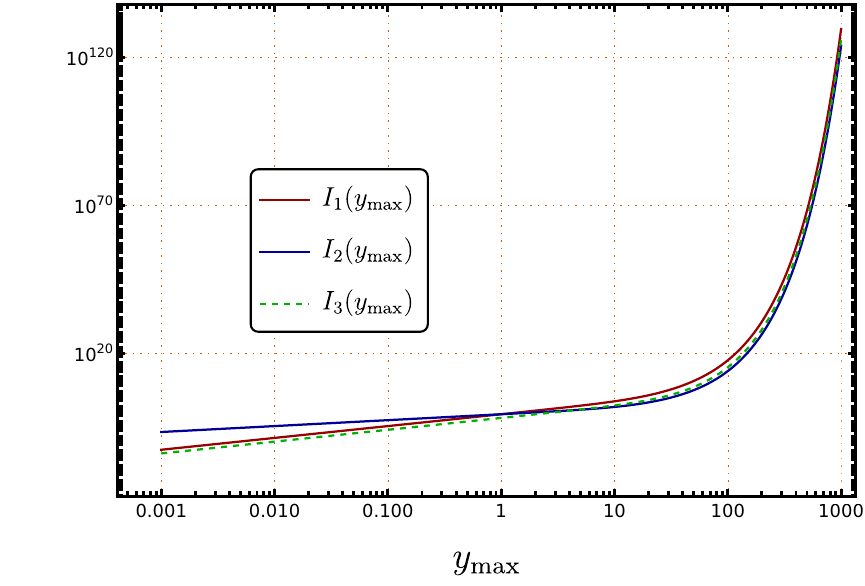}
	\hspace{.2cm}
	\includegraphics[width=0.48\textwidth]{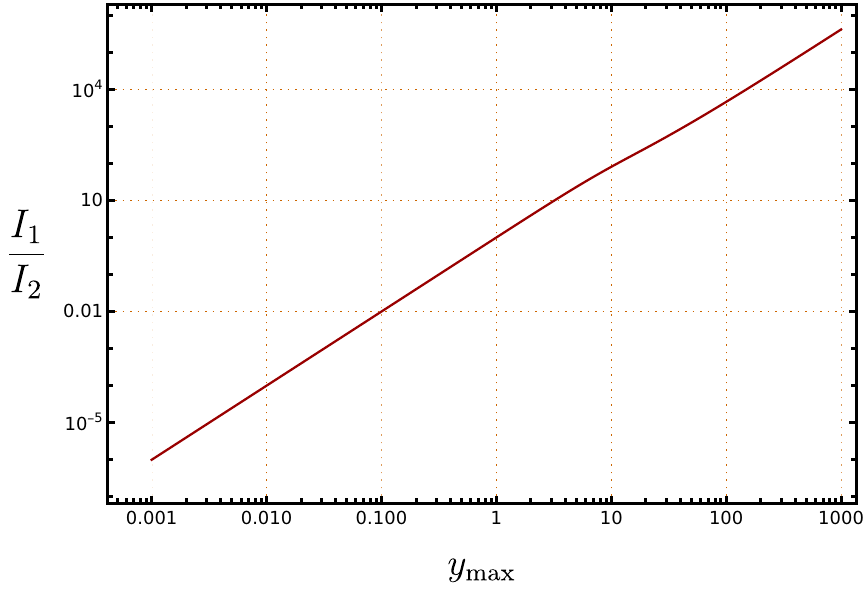}
	\caption{%\scriptsize 
		\textbf{Left:} Numerical calculation of \eqref{I1}, \eqref{I2}, and \eqref{I3}. \textbf{Right:} Comparison of \eqref{I1} and \eqref{I2} in term of $y_{\rm max}$. As seen for a wide range of $y_{\rm max}$, the expression of \eqref{C-2} is good enough for the regime of $R \gg 10^4$.}
	\label{fig:I1I2_and_I1OvereI2}
}
%%%%%%%%%%%%%%%%%%%%%%%%%%%%%%%%%%%%%%%%%
\fg{
	\centering
	\includegraphics[width=0.48\textwidth]{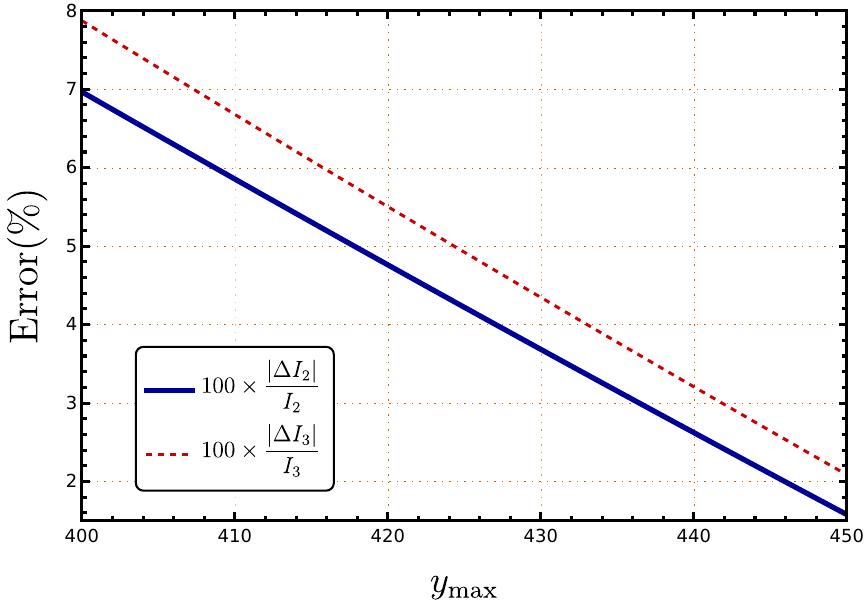}
	\caption{%\scriptsize 
		The relative errors for the estimations of \eqref{I2-estimation}-\eqref{I3-estimation} with respect to the \eqref{I2}-\eqref{I3} for large values of $y_{\rm max}$. As seen, in the range of our interest the relative errors are less than $10\%$.}
	\label{fig:errors}
}
%%%%%%%%%%%%%%%%%%%%%%%%%%%%%%%%%%%%%%%%

With \eqref{rho} and the definition in \eqref{rho-param}, we can express the parameter ${\cal C}$ as
\begin{align}
\label{C}
{\cal C} 
%= \dfrac{\rho^{(A)}_c}{\rho_c} 
&= 
\dfrac{I_2(y_{\rm max})}{12\pi^2}R\left(
\dfrac{H_c}{\Mpl}
\right)^2\Bigg(1+
\dfrac{1}{R}\dfrac{I_1(y_{\rm max})}{I_2(y_{\rm max})}
\Bigg)\,,
\end{align}
where in the last equality we have used \eqref{mA}. As seen in the right panel of Fig.~\ref{fig:I1I2_and_I1OvereI2}, in the cases $R \gg 10^4$ which we will show is the case here, one can safely neglect the second term above for a broad range of $y_{\rm max}$ and estimate \eqref{C} as
\begin{align}
\label{C-2}
{\cal C} &\simeq 
%\dfrac{R}{12\pi^2}\left(\dfrac{H_{\rm c}}{\Mpl}\right)^2~I_2(y_{\rm max}) =
\dfrac{\e^2\sqrt{6}}{30\pi^2\sqrt{\lambda}}\left(\dfrac{f_{\rm EDE}}{8\%}
\right)^{1/2}
\left(
\dfrac{H_c}{\Mpl}
\right)~I_2(y_{\rm max}) \,,
\end{align}
where we have used \eqref{R_2}.

In the limit $y_{\rm max} \ll 1$, we can use the small argument limits of Bessel functions and calculate the integral \eqref{I2} analytically with a good accuracy. In this limit, we obtain
\begin{align}
{\cal C} \simeq  \dfrac{ 
	y_{\rm max}^2}{48\pi^2 }\bigg(
\dfrac{m_A}{\Mpl}
\bigg)^2 = \dfrac{ 
	y_{\rm max}^2}{48\pi^2 }R\left(
\dfrac{H_c}{\Mpl}
\right)^2 \,.
\end{align}
From \eqref{R_2}, we observe that $R\left(H_c/\Mpl\right)^2 \propto \left(H_c/\Mpl\right)$. Moreover, since during the RD era $H \ll \Mpl$, there is no room for ${\cal C}\sim {\cal O}(1)$.

In the limit $y_{\rm max} \gg 1$, the integrands exhibit behavior resembling a delta function centered around $y_{\rm max}/\pi$. Consequently, within the range $100<y_{\rm max}<700$, we can approximate the integrals \eqref{I1} and \eqref{I3} as follows:
\begin{align}
\label{I2-estimation}
I_2(y_{\rm max}) &\simeq \dfrac{y_{\rm max}^2}{2\pi^4}~e^{s\, y_{\rm max}} \,,
%\hspace{1cm}
%s = \dfrac{4}{5}\left(\pi^{-1/2}-\pi^{-4/3}\right) \,,
\\
\label{I3-estimation}
I_3(y_{\rm max}) &\simeq \dfrac{2y_{\rm max}}{\pi^2}I_2(y_{\rm max}) \,.
\end{align}
where $s \simeq 0.3$~. The relative errors with respect to the original integrals \eqref{I1} and \eqref{I3} are displayed in Fig.~\ref{fig:errors}, covering the range of interest for $y_{\rm max}$.

Knowing the value of ${\cal C}$ enables us to calculate the relic dark photon energy density, which may constitute either a part or the entirety of the dark matter energy density. We are interested in the parameter space where the dark photon becomes non-relativistic immediately after symmetry breaking, characterized by $R \gg 1$. 

Using \eqref{rho-e} and the conservation of entropy density, $g_{*s{\rm c}} a_{\rm c}^3 T_{\rm c}^3=g_{*s0} a_0^3 T_0^3$, we can estimate the Hubble parameter at $z_{\rm c}$ to be
\begin{align}
	\left(
	\dfrac{H_{\rm c}}{\Mpl}
	\right)^2 &
	=\dfrac{\pi^2}{90}g_{*{\rm c}}\left(
	\dfrac{g_{*s{\rm c}}}{g_{*s0}}
	\right)^{4/3}\left(
	\dfrac{a_0}{a_{\rm c}}
	\right)^4\left(
	\dfrac{T_0}{\Mpl}
	\right)^4
	\nonumber\\
	&\simeq 2.84 \times 10^{-112}  
	\left(1+\dfrac{z_{\rm eq}}{z_{\rm c}}\right)
	\left(
	\dfrac{z_{\rm c}}{5000}
	\right)^4 \,,
	\label{H_MP}
\end{align}
where we have considered $T_0 \simeq 10^{-13} {\rm GeV}$. Therefore, the parameter $R$, defined in \eqref{R_2}, is given by
\begin{align}
	\label{R_3}
	R \simeq 5.85 \times 10^{55} \dfrac{\e^2}{\sqrt{\lambda}}\left(
	\dfrac{f_{\rm EDE}}{8 \%}
	\right)^{1/2}\left(1+\dfrac{z_{\rm eq}}{z_{\rm c}}\right)^{-1/2}
	\left(
	\dfrac{z_{\rm c}}{5000}
	\right)^{-2} \,.
\end{align}
This means that for typical values for $\e$ and $\lambda$, parameter $R$ is very very large in such a way that after the end of EDE phase, the dark photon becomes massive and it never experiences relativistic phases. Since the produced dark photons do not contribute to the radiation components after their generation, there is no constraint on ${\cal C}$ from the relativistic degree of freedom during RD era. We define the fractional energy density of dark photons at the present time as:
\begin{align}
	\Omega_{A} \equiv \dfrac{\rho^{(A)}_0}{\rho_0} &= \dfrac{\rho^{(A)}_0}{\Omega_{\rm m}^{-1}\rho_{\rm m,0}}= \Omega_{\rm m} \dfrac{\rho^{(A)}(z_{\rm c})}{\rho^{\rm (m)}(z_{\rm c})}
	\label{Omega_A}
\end{align}
where $\Omega_{\rm m} \equiv \rho_{\rm m,0}/\rho_{0} $ is the fractional density of matter today and $\rho^{\rm (m)}(z_{\rm c})$ is the total energy density of matter at $z_c$. We can consider that the matter component at that time consists of two parts: $\rho^{\rm (m)}(z_{\rm c})= \rho^{\rm (A)}(z_{\rm c})+\rho^{\rm (D)}(z_{\rm c})$; the non-dark photon part $\rho^{\rm (D)}(z_{\rm c}) \equiv {\cal D} \rho(z_{\rm c})$ and the dark photon part $\rho^{(A)}(z_{\rm c})$. In this case, the fractional energy density of the dark photon to the total dark matter is given by:
\begin{align}
	\label{fA}
	f_A &\equiv \dfrac{\Omega_{A}}{\Omega_{\rm m}} = 
	\dfrac{\cal C}{\cal C+D}\,.
\end{align}
The energy density of dark photons after generation %, denoted as $\rho^{(A)}_c$, 
constitutes only a small fraction of the total energy density at that time. %$\rho_{\rm c}$. 
This arises from several factors. Firstly, the vacuum energy of the waterfall field is dominant in comparison to the axion's energy, as assumed in \eqref{rho_axion}, although it remains subdominant when compared with the total energy density \eqref{rho_psi}. Subsequently, the dark photons are generated as a result of axion's slow-rolling (before $z_c$) and acquire mass by extracting vacuum energy from the potential \eqref{pot} (after $z_c$). Secondly, a portion of the energy is used in the formation of cosmic strings at the end of the waterfall symmetry breaking~\cite{Felder:2000hj,Felder:2001kt}\footnote{The vacuum energy of the waterfall field is indeed released as the field undergoes symmetry breaking. It is initially transformed into the kinetic and potential energy of the field itself and its excitations. It can produce various particles, including scalar and gauge bosons, which can carry a portion of the released energy. This is often a non-equilibrium process and can lead to the production of particles with significant momenta. Cosmic strings are formed as a result of the symmetry-breaking phase transition. The energy stored in the gradients and misalignment of the field configuration in the vicinity of cosmic strings is what contributes to their energy density. This energy is not directly transferred from the vacuum energy of the waterfall field but is a consequence of the change in the field's configuration during the phase transition.}. In essence, if we neglect possible dissipation and ignore the formation of cosmic strings, the continuity equation for the energy density at $z_{\rm c}$ %$z_{\rm c}^-$ and $z_{\rm c}^+$ 
implies that $\rho^{\rm (A)}(z_{\rm c}) \leqslant \rho^{(\psi,\phi)}(z_{\rm c}) \simeq f_{\rm EDE} \, \rho(z_{\rm c})$. This results in the following constraint:
\begin{align}
	\label{eq:C_fEDE}
	{\cal C} \lesssim f_{\rm EDE} \,. %\sim 8\%
\end{align}
Considering ${\cal D + C} \simeq z_{\rm eq}/(z_{\rm eq}+z_{\rm c})$ (see App.~\ref{C_upper_bound} for more details), the relation of \eqref{fA} leads to
\begin{align}
	f_A  \lesssim {\cal C} \left(
	1+\dfrac{z_{\rm c}}{z_{\rm eq}} 
	\right) \,.
	\label{f_02}
\end{align}
Since our goal is to minimize the modification to the well-established history of the universe, we set the location of MR equality to the well-known value of $z_{\text{eq}} \simeq 3400$. Depending on the location of the end of symmetry breaking, dark photons either make up a fraction or the entirety of DM.  In what follows, we consider two cases $z_c \sim z_{\rm eq}$ and $z_c \gg z_{\rm eq}$ and investigate the allowed values for the fraction $f_A$.

\begin{itemize}
	\item $\boldsymbol{z_{\rm c} \sim z_{\rm eq}}$: \\
	If the waterfall field becomes tachyonic around the MR equality, an upper limit on the dark photon fraction within DM can be established, typically at the order of $f_A \sim {\cal O}(10\%)$. This assumes that the parameter ${\cal C} \sim {\cal O}(5\%)$, a choice made to address the $H_0$ tension. Consequently, referring to equation \eqref{C+D}, it follows that the generated dark photons cannot account for the majority of dark matter, as indicated by the non-zero value of ${\cal D}$.
	
	\item $\boldsymbol{z_{\rm c} \gg z_{\rm eq}}$:\\
	The waterfall field becomes tachyonic well before MR equality. Therefore, if the dark photon production ends appropriately, which we will demonstrate for a wide range of model parameters, it's possible to model the total present DM using the dark photons generated during RD. To be precise, by setting $f_A = 1$, we determine that the symmetry breaking ends at
	\begin{align}
		\label{zc_DM}
		z_{\rm c} = z_{\rm eq} \left({\cal C}^{-1}-1\right) %\simeq \dfrac{z_{\rm eq}}{{\cal C}} 
		\simeq \dfrac{z_{\rm eq}}{{\cal C}} \sim {\cal O}(10^4) \,,
	\end{align} 
	where we have used \eqref{eq:C_fEDE}. 
	While it is important to acknowledge that this scenario might not completely resolve the $H_0$ tension, conducting a comprehensive Markov Chain Monte Calro (MCMC) simulation is essential to arrive at a definitive conclusion. Nonetheless, there is potential to alleviate the tension at significantly high redshifts. For instance, as shown in Ref.\cite{Poulin:2018cxd}, a radiation-like EDE cosmology with the mean value of $z_{\rm c} \sim 10^4$ exhibits a preference relative to $\Lambda$CDM.
\end{itemize}

In the next section, we will explore the parameter space of the model to ensure its consistency and address the Hubble tension, as well as the origin of DM.

\section{$H_0$ tension and/or dark photon DM}
\label{sec:EDE}

In the preceding section, we investigated the generation of dark photon arising from the rolling of the axion field. Following that, we estimated the cumulative energy density of dark photons once they become massive at the end of the waterfall symmetry breaking. These dark photons can provide an explanation for the origin of DM. Importantly, the use of the waterfall mechanism allows for the termination of EDE phase without the need for intricate potentials or fine-tuning. Consequently, this setup has the potential for a successful EDE phase that could relax the $H_0$ tension. In this section, our aim is to identify a parameter space that allows for the realization of the waterfall symmetry breaking, addressing both the $H_0$ tension and the origin of DM, a fraction of it or its entirely.

For a comprehensive and precise analysis, it is imperative to take into account two critical aspects. Firstly, the nonlinear dynamics of the waterfall field at the end of the symmetry breaking must be carefully considered. Secondly, conducting a MCMC simulation is essential to determine the allowed parameter space accurately. In the following, we intend to overcome these two challenges. 
%and streamline these issues to the greatest extent possible.

Spontaneous symmetry breaking is a strongly nonlinear effect. Perturbative methods are inadequate for describing the formation of cosmic strings and the interaction of particles produced by tachyonic instability. Fortunately, numerical simulations using methods such as lattice simulations have been employed to address these complexities (see \cite{Felder:2000hq}). The symmetry breaking phase is accompanied by a release of energy. This energy is stored in the gradients and 
%misalignment 
the vev of the field within and around the cosmic string. %The energy density associated with these defects is what constitutes the energy of the cosmic strings. 
As demonstrated in \cite{Felder:2000hj}, these instabilities rapidly convert most of the initial potential energy density $\rho_{\rm EDE}$ into various forms of energy, including the mass of dark photons, the energy of produced particles, and the energy associated with field gradients. 
%This transformation of energy is a key factor contributing to the formation and properties of cosmic strings. The energy stored in the gradients and misalignment of the field configuration in the vicinity of cosmic strings is what contributes to their energy density. 
Considering \eqref{string-mu}, one can show that
\begin{align}
	G \mu \simeq \dfrac{3\e^2}{2\pi \lambda}\dfrac{f_{\rm EDE}}{R} \lll 1 \,,
\end{align}
which totally solves concerns about cosmic string production at the end of EDE phase. %%%%%%%%%%%%%%%%%%%%%%%%%%%%%%%%%%%%%%%%%%%%%%%%%%%%%%%%%%%%%%%%%%%%%%%%%%%%%%%%%%%%%%%%%%%%%%%%%%%%%%%%%%%%%%%%%%%%%%%%%{\color{red}Comparison between energy of the dark photons and cosmic string (gradients energy)... plots!!!???}

Since the potential energy of the waterfall field is instantaneously converted into dark photons~\cite{Salehian:2020asa,Nakagawa:2022knn} and cosmic strings~\cite{Felder:2000hj,Felder:2001kt}, we can avoid the technicalities of taking the waterfall field dynamics into account~\cite{Abolhasani:2010kr,Gong:2010zf}. %\textcolor{red}{
	Base on \cite{Felder:2000hj}, in order to tachyonic instability and fields relax near the completion of symmetry breaking, we need $g^2 \gg \lambda$.
	%Base on \cite{Felder:2000hj}, in order to tachyonic instability and fields relax near the minimum of the potential within a single oscillation, we need $g^2 \gg \lambda$.
%} {\color{red} 
Additionally, we make the assumption that a significant portion of the energy is transferred to dark photons rather than cosmic strings.
%}  
During the RD era, cosmic strings are diluted as like radiation. Consequently, in our scenario, the EDE component also dilutes like radiation\footnote{Note that the notation of $n=2$ as used in \cite{Poulin:2018cxd} corresponds to $n=4$ in our framework.}.

According to Ref.~\cite{Poulin:2018cxd}, the mean and best-fit parameters $\{f_{\rm EDE}, z_{\rm c}\}$ estimated by the MCMC simulation for radiation-like EDE cosmology are given by $\{2.8\%, 13676\}$ and $\{4.4\%, 5345\}$, respectively. This model exhibits a slight preference over $\Lambda$CDM, as it reduces the $\chi^2$ by -9.5\,. Although a MCMC search of the parameter space is imperative to demonstrate the success of a EDE cosmology, we adopt their best-fit parameters as reference values for our EDE scenario, because the cosmological evolution of EDE in our scenario is almost the same as theirs except for new DM component after $z_{\rm c}$. A similar approach was employed in \cite{Nakagawa:2022knn}, wherein a dark Higgs is trapped at the origin for a long time, thus realizing an EDE scenario. We will leave a more comprehensive MCMC analysis to future works.
%In \cite{Nakagawa:2022knn}, the authors have considered a dark Higgs trapped at the origin using the dark photons non-thermally generated by coherent oscillations of axion. The dark Higgs is trapped for a long time, which realizes EDE scenario, and after the end of trapping around the matter-radiation equality, the dark Higgs quickly decays into the dark photon. Interestingly, this scenario not only is free from fine-tuning problem, but also the axion fields can explain all DM. This scenario for EDE seems to be almost the same as~\cite{Poulin:2018cxd} except for oscillatory features at $z < z_c$. Therefore, Nakagawa et.al. \cite{Nakagawa:2022knn} have adopted the best-fitparameters estimated by the Markov Chain Monte Calro (MCMC) simulation of \cite{Poulin:2018cxd}, i.e. $\{5\%, 5000,4.5\}$, as reference values for their EDE scenario.

Taking into account the best-fit $\{f_{\rm EDE}, z_{\rm c}\}=\{4.4\%, 5345\}$, we subsequently normalize the relations by dividing by $f_{\rm EDE}=5\%$ and $z_{\rm c}=5000$ for convenience.
%for the sole purpose of enhancing the clarity and visual appeal of the expressions. 
Therefore, the Hubble parameter $H_{\rm c}$, the mass of dark photons $m_A$, $M$ and $\phi_{\rm c}$ are given by
\begin{align}
H_{\rm c} &\simeq 4 \times 10^{-29}~ {\rm eV} \left(1+\dfrac{z_{\rm eq}}{z_{\rm c}}\right)^{1/2}\left(
\dfrac{z_{\rm c}}{5000}
\right)^{2} 
\\
m_A &\simeq 0.3~ {\rm eV}~ \dfrac{\e}{\lambda^{1/4}}\left(
\dfrac{f_{\rm EDE}}{5 \%}
\right)^{1/4}\left(
\dfrac{z_{\rm c}}{5000}
\right)\left(1+\dfrac{z_{\rm eq}}{z_{\rm c}}\right)^{1/4}
\\
M &\simeq 0.3~ {\rm eV}~ \lambda^{1/4}\left(
\dfrac{f_{\rm EDE}}{5 \%}
\right)^{1/4}\left(
\dfrac{z_{\rm c}}{5000}
\right)\left(1+\dfrac{z_{\rm eq}}{z_{\rm c}}\right)^{1/4}
\\
\label{phic}
\phi_{\rm c} &\simeq 0.3~ {\rm eV}~\left(
\dfrac{\lambda^{1/4}}{g}
\right) \left(
\dfrac{f_{\rm EDE}}{5 \%}
\right)^{1/4}\left(
\dfrac{z_{\rm c}}{5000}
\right)\left(1+\dfrac{z_{\rm eq}}{z_{\rm c}}\right)^{1/4}
\end{align}
%Combining \eqref{H_MP} and \eqref{C-2}, we obtain
%\begin{align}
%\label{C-3}
%{\cal C} &\simeq
%1.38 \times 10^{-58}~I_2(y_{\rm max})~ \dfrac{\e^2}{\sqrt{\lambda}}\left(
%\dfrac{f_{\rm EDE}}{8 \%}
%\right)^{1/2}\left( 1+\dfrac{z_{\rm eq}}{z_{\rm c}}\right)^{1/2}\left(
%\dfrac{z_{\rm c}}{5000}
%\right)^{2}
%\end{align}
We are interested in the fractional density of the dark photon to the dark matter at the present which, combining \eqref{H_MP}, \eqref{C-2}, and \eqref{fA}, is given by
\begin{align}
	\label{fA_EDE}
	f_A &\simeq 10^{-58}~I_2(y_{\rm max})~ \dfrac{\e^2}{\sqrt{\lambda}}\left(
	\dfrac{f_{\rm EDE}}{5 \%}
	\right)^{1/2}\left( 1+\dfrac{z_{\rm eq}}{z_{\rm c}}\right)^{1/2}\left(
	1+\dfrac{z_{\rm c}}{z_{\rm eq}}
	\right)\left(
	\dfrac{z_{\rm c}}{5000}
	\right)^{2} \,.
\end{align}
To estimate the back-reaction, it is also convenient to rewrite \eqref{S} as
\begin{align}
	\label{S2}
	S_\phi &\simeq 2.7 \times 10^{-59}~I_3(y_{\rm max})\bigg(
	\dfrac{\alpha}{f_a}{\rm eV}
	\bigg)^2\left(1+\dfrac{z_{\rm eq}}{z_{\rm c}}\right)
	\left(
	\dfrac{z_{\rm c}}{5000}
	\right)^4 \,.
\end{align}
In Fig.~\ref{fig:fAS}, the fractional density of the dark photon and back-reaction on axion field has been plotted as $\lambda^{1/4} \sim \e$. We have used the best-fit reported by \cite{Poulin:2018cxd} for radiation-like EDE component and looked for the allowed parameter space respecting $S_\phi <0.1$. We found that the regime $441<y_{\rm max}<446$ for axion coupling $\alpha/f_a \lesssim 0.1~ {\rm eV}^{-1}$ can make up a non-negligible fraction (up to $f_A \lesssim 0.2$) of the observed DM.

As previously mentioned, for the dark photons to have the potential to account for the entirety of DM, the waterfall symmetry breaking must occur at a higher redshift. According to equation \eqref{f_02}, a suitable case is $z_{\rm c} \simeq 4\times 10^{4}$. While this value may not help resolving the $H_0$ tension, we consider this case in the subsequent analysis to address the origin of DM. Under this regime, relations \eqref{fA_EDE} and \eqref{S2} can be reformulated as follows
\begin{align}
	f_A &\simeq 4.6 \times 10^{-55}~I_2(y_{\rm max})~ \dfrac{\e^2}{\sqrt{\lambda}}\left(
	\dfrac{f_{\rm EDE}}{5 \%}
	\right)^{1/2}\left(
	1+\dfrac{z_{\rm c}}{z_{\rm eq}}
	\right)\left(
	\dfrac{z_{\rm c}}{4\times 10^4}
	\right)^4 \,,
	\label{f_A-DM}
	\\
	\label{S2-DM}
	S_\phi &\simeq 1.2 \times 10^{-55}~I_3(y_{\rm max})\bigg(
	\dfrac{\alpha}{f_a}{\rm eV}
	\bigg)^2
	\left(
	\dfrac{z_{\rm c}}{4\times 10^4}
	\right)^4 \,.
\end{align}
We have also depicted this region in Fig.~\ref{fig:fAS} and observed that the parameter range $409<y_{\rm max}<417$ by assuming $\lambda^{1/4} \sim \e$ and the axion coupling $\alpha/f_a \lesssim 0.3~ {\rm eV}^{-1}$ has the potential to account for a significant portion of the observed DM content.

In our calculations, we have identified a significant dependence of the fractional energy density of the dark photon $f_A$ on the parameter $y_{\rm max}$. In addition, the back-reaction $S_\phi$ depends on both parameters  $y_{\rm max}$ and the coupling $\frac{\alpha}{f_a}$. Notably, according to \eqref{ymax}, we have established that $y_{\rm max}$ is a function of $\frac{\alpha}{f_a}$, introducing an interdependence between these key variables. To enhance the generality and applicability of our findings, it is prudent to re-examine our calculations in terms of its free parameters. 
%${\phi_0}, \lambda, g, \alpha, f_a, \cdots$. 
This approach allows for a more flexible exploration of the parameter space and facilitates a clearer understanding of the implications of different choices for the free parameters on the final results. By emphasizing the relation between $y_{\rm max}$ and $\frac{\alpha}{f_a}$, we can provide a more comprehensive analysis, ultimately enhancing the robustness of our model.
%%%%%%%%%%%
\fg{
	\centering
	\includegraphics[width=0.48\textwidth]{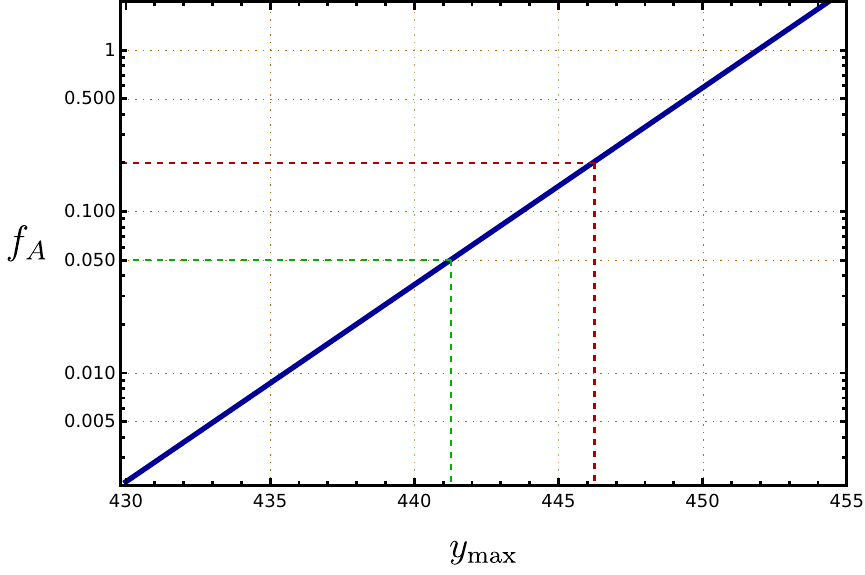}
	\includegraphics[width=0.48\textwidth]{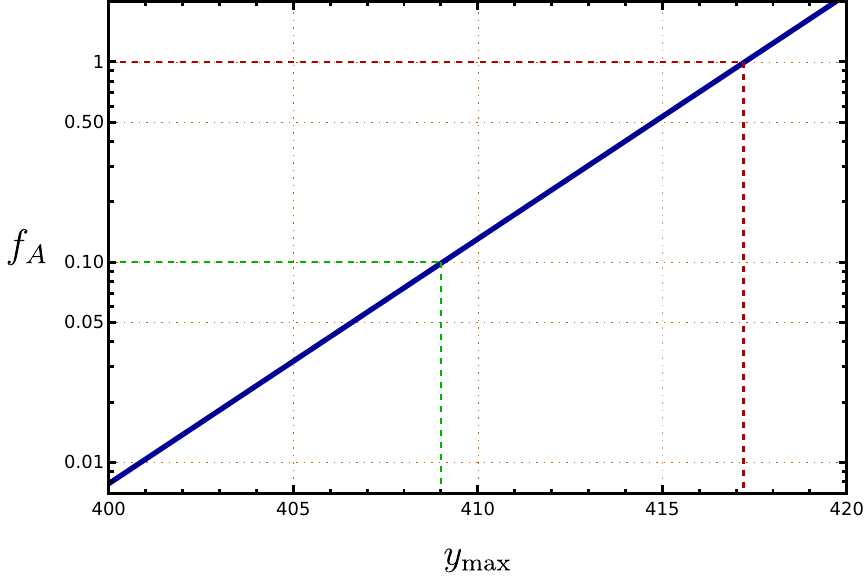}
	\\
	\vspace{.3cm}
	\includegraphics[width=0.48\textwidth]{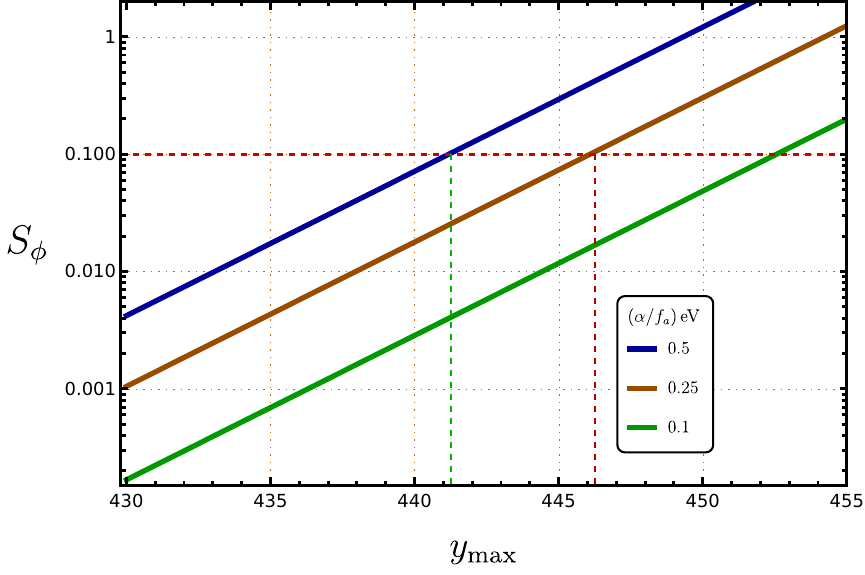}
	\includegraphics[width=0.48\textwidth]{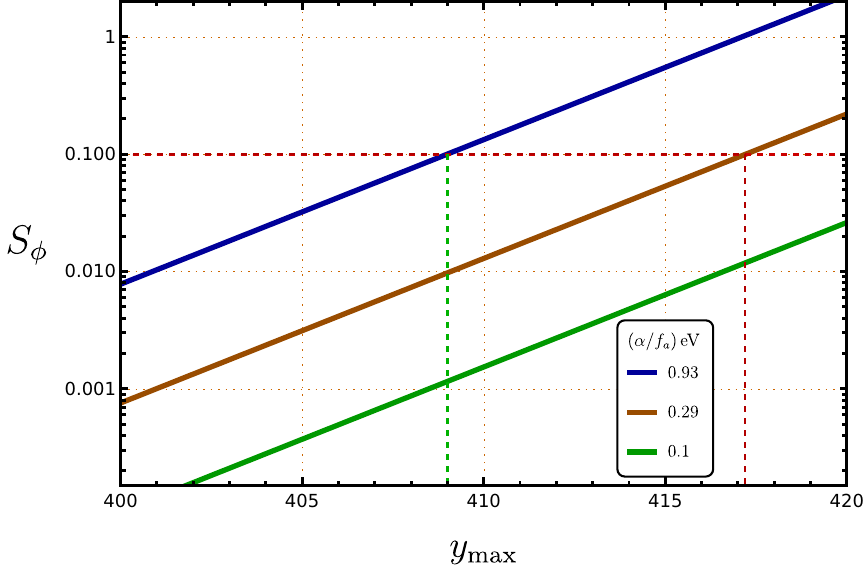}
	\caption{%\scriptsize  
		We have plotted the fractional density of the dark photon $f_A$ and it's back-reaction on axion field $S_\phi$ \eqref{S2} in terms of $y_{\rm max}$. It was assumed  $\lambda^{1/4} \sim \e$ to calculate $f_A$. We fixed $z_{\rm eq}=3400$ while looking for the small back-reaction regime, i.e. $S_\phi<0.1$.  (\textbf{Left plots: $z_{\rm c}=5000$}) We are interested in the range $441<y_{\rm max}<446$ for axion coupling $\left(		\alpha/f_a \right){\rm eV} \lesssim 0.1$ where it propose a successful EDE cosmology and makes a portion of DM $0.05<f_A \leqslant 0.2$. (\textbf{Right plots: $z_{\rm c}=4\times10^4$}) The axion coupling $\left(		\alpha/f_a \right){\rm eV} \lesssim 0.3$ for the range $409<y_{\rm max}<417$ can lead to  DM is generated significantly $0.1<f_A \leqslant 1$.}
	\label{fig:fAS}
}
%%%%%%%%%%%%%%%%%%%%%%%%%%%%%%%%%%%%%%%%%%

To delve further, we consider the dark photon becomes massive instantaneously at $t=t_{\rm c}$ and use the definition of \eqref{ymax} to find
\begin{align}
\label{ymax2}
y_{\rm max} &\equiv \dfrac{1}{5}\dfrac{\alpha}{f_a}\phi_0 \Big(
\dfrac{m}{H_{\rm c}}
\Big)^{2} \,.
\end{align}
Combining with \eqref{phi_cc} and \eqref{phic}, we obtain
\begin{align}
y_{\rm max} = 1.2 \left(
\dfrac{\phi_0}{\phi_{\rm c}}-1
\right)\left(
\dfrac{\alpha}{f_a}{\rm eV}
\right)
\left(
\dfrac{\lambda^{1/4}}{g}
\right)\left(
\dfrac{f_{\rm EDE}}{8 \%}
\right)^{1/4}\left(
\dfrac{z_{\rm c}}{5000}
\right) \left(1+\dfrac{z_{\rm eq}}{z_{\rm c}}\right)^{1/4} 
\end{align}
The above parametrization shows that the parameter $y_{\rm max}$ is a function of free parameters of the model as $y_{\rm max}\left({\tiny 
	\frac{\phi_0}{\phi_{\rm c}},\frac{\lambda^{1/4}}{g},\frac{\alpha}{f_a}{\rm eV}}
\right)$. 
In addition, from \eqref{phi0phic}, we find the allowed parameter space for the axion field as $0<\frac{\phi_0}{\phi_{\rm c}}-1 < 0.25$.

%\begin{align}
%\label{phi0phic-1}
%0<\dfrac{\phi_0}{\phi_{\rm c}}-1 < 0.25
%\end{align}

Considering a fixed value for the parameters $f_{\rm EDE}, z_c$, and $z_{\rm eq}$, %, and assuming $\lambda^{1/4} \sim \e$ 
we find free parameters to be $\frac{\alpha}{f_a},\frac{\lambda^{1/4}}{g}$, and $\frac{\phi_0}{\phi_{\rm c}}$. In Fig.~\ref{fig:fA_alpha_g}, we presented the 3D plots illustrating the behavior of $f_A\left(
\frac{\alpha}{f_a},\frac{\lambda^{1/4}}{g}
\right)$ concerning different values of $\frac{\phi_0}{\phi_{\rm c}}$ for two distinct scenarios: (i) an EDE model aimed at resolving the $H_0$ tension, and (ii) a dark photon dark matter model targeting the origin of DM. Initial observations, without conducting an MCMC search across the parameter space, suggest that our current framework may not simultaneously address both the $H_0$ tension and the complete origin of DM. At this stage, a more comprehensive assessment through an MCMC analysis is deferred to future studies. This thorough exploration of the parameter space is crucial to ascertain the model's capacity to reconcile both cosmological challenges ($H_0$ tension and the origin of DM) simultaneously.

%%%%%%%%%%%%%%%%%%%%%%%%%%%%%%%%%%%%%%%%%%%%%%%%%%%%%%% How I plot Fig.6 %%%%%%%
\iffalse
Demanding $0.05 \leqslant f_A \leqslant 0.2$ leads to
\begin{align}
y_1 = 441.25 \leqslant y_{\rm max} \leqslant y_2= 446.25
\end{align}
imposing this interval in $I_3$ and demanding $S < 0.1$, we can find the  maximum value of $\frac{\alpha}{f_a}{\rm eV}$ using \eqref{S2} or \eqref{S2-DM} for each $y_{\rm max}$ in the interested range
which is a function of $y_{\rm max}$. One can put a constraint on $\lambda^{1/4}/g$ when we use \eqref{phi0phic}. The results are in Fig.~\ref{fig:fA_alpha_g}.
\fi
%%%%%%%%%%%%%%%%%%%%%%%%%%%%%%%%%%%%%%%%%%%%%%%%%%%%%%%%%%%%%%%%%%%%%%%%%%%%%
\iffalse
\section{Parameter Space}

Demanding $z_{\rm eq}$ to be a fixed value, our model have 8-dim parameter space. Free parameters include
\begin{align}
	\{ \e, ~\lambda, ~m, ~g, ~\phi_0, ~\frac{\alpha}{f_a}, ~f_{\rm EDE},~ z_c\}
\end{align}
Note that having $z_c$ identifies $H_c$ and combining with $\lambda$ and $f_{\rm EDE}$, one can fix $M$.

A lot of parameter space with respect to EDE with $\{f_{\rm EDE}, z_c, n \}$!!! we have 6 parameters more than previous EDE models!!!

\textcolor{red}{Base on \cite{Felder:2000hj}, in order to tachyonic instability and fields relax near the minimum of the potential within a single oscillation, we need $g^2 \gg \lambda$.}

Note that since $g^2 \psi_{\rm min}^2 \gg H_{\rm c}^2$, by assuming $\lambda^{1/4} \sim \e$, we get to
\begin{align}
	\dfrac{\lambda^{1/4}}{g} \sim \dfrac{\e}{g} \ll \sqrt{R} \sim 10^{27}
\end{align}
\fi

%%%%%%%%%%%
\fg{
	\centering
	\includegraphics[width=0.8\textwidth]{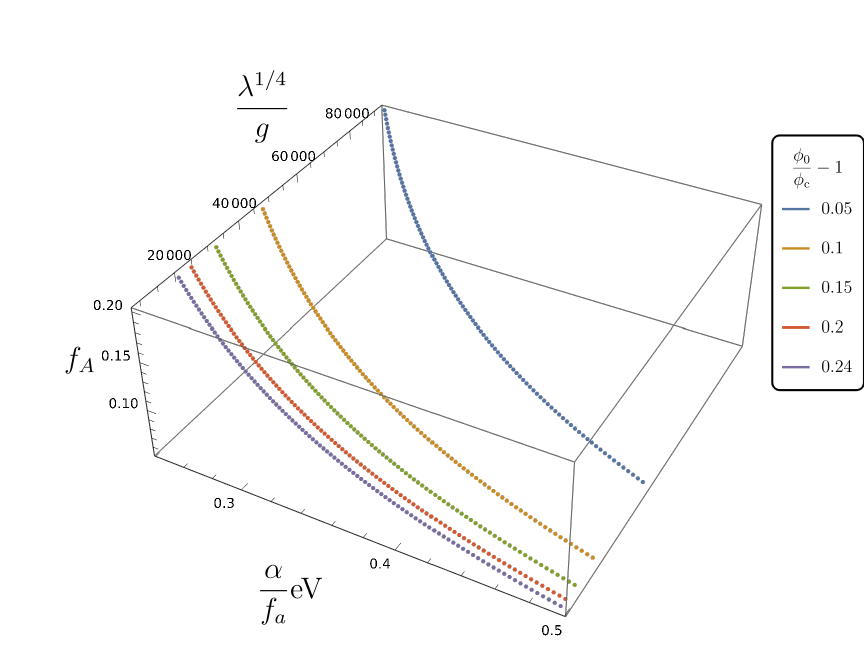}
	\\
	\includegraphics[width=0.8\textwidth]{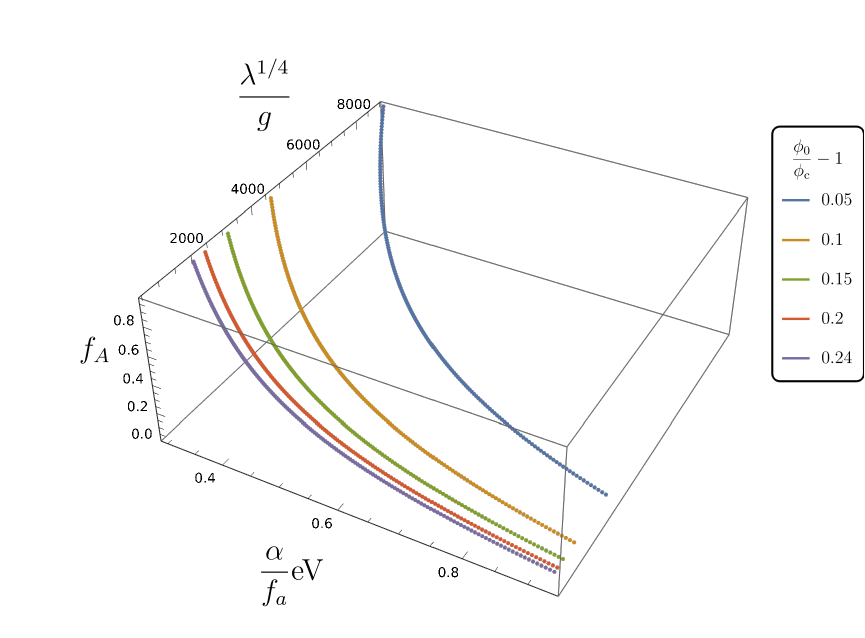}
	\caption{%\scriptsize 
		3D plots for $f_A\left(
		\frac{\alpha}{f_a},\frac{\lambda^{1/4}}{g}
		\right)$ for various values $\frac{\phi_0}{\phi_{\rm c}}$ %as $\lambda^{1/4} \sim \e$ 
		for two scenarios: (\textbf{Top:}) a successful EDE model to alleviate $H_0$ tension , %\eqref{f_A-DM}
		 and (\textbf{Bottom:}) a dark photon dark matter model  to address the origin of DM.}
	\label{fig:fA_alpha_g}
}
%%%%%%%%%%%

%%%%%%%%%%%%%%%%%%%%%%%%%%%%%%%%%%%%%%%%%%
\section{Summary and Discussion}
\label{sec:summary}
Within the standard model of cosmology, significant attention has been directed towards two pivotal questions: the $H_0$ tension and the origin of dark matter component. %Various models have been proposed as the origins of dark matter and potential resolutions for the $H_0$ tension \cite{Kamionkowski:2022pkx, Moshafi:2022mva, Poulin:2018cxd, Poulin:2018dzj, Smith:2019ihp, Niedermann:2019olb, Niedermann:2021vgd}. 
In this paper we have proposed a novel scenario in which the existence of a waterfall symmetry breaking during the radiation-dominated era can address these two issues. The model comprises a complex waterfall field, an axion field, and the gauge field (dark photon). Before symmetry breaking, the waterfall field contributes to the total energy density with a constant term, establishing a novel scenario for the EDE proposal. Meanwhile, dark photon are generated non-perturbatively due to the tachyonic instability coming from the Chern-Simons interaction with the axion field. 

When the axion field reaches the given critical value, the waterfall field becomes tachyonic, initiating a rapid roll towards its global minimum. Under the assumption of a very heavy waterfall mass, the transition to the global minimum and symmetry breaking occurs promptly. More interestingly, after the waterfall becomes tachyonic at the end of EDE, the dark photon acquires mass contributing to the dark matter abundance. 

The breaking of the symmetry is accompanied by the release of energy into various components, including the mass of dark photons and the generated cosmic string. We have assumed that a significant portion of the energy is transferred to dark photons.  During the RD era, the cosmic string energy density is diluted as like radiation. Consequently, in our scenario, the remnant EDE component dilutes like radiation. Therefore, we have safely adopted the best-fit parameters of \cite{Poulin:2018cxd} for a successful radiation-like EDE scenario.

We have identified viable model parameters where the Hubble tension is resolved by EDE, and the dark photons elucidate a substantial portion of the observed DM across a wide parameter space. Additionally, we have delineated a parameter space in which dark photons constitute the entirety of the observed DM.

While our semi-analytical analysis highlights the model's potential to address each issue separately, it becomes evident that our current setup cannot resolve them simultaneously. Conducting a comprehensive lattice simulation and a series of MCMC runs is crucial for a more nuanced understanding. We defer these simulations to future studies, aspiring to explore a broader parameter space that encompasses the complexity of the symmetry-breaking mechanism.

Contemplating the core concept of this work, one might consider the simplest approach to tackle both the Hubble tension and the origin of DM concurrently: employing double waterfall fields. The initial symmetry breaking takes place at a very high redshift, followed by a secondary waterfall transition closer to MR equality. This setup necessitates a larger parameter space. The current version involves an 8-dimensional parameter space. Requiring $z_{\rm eq}$ to remain fixed, the free parameters encompass\footnote{Having $z_c$ identifies $H_c$ and combining with $\lambda$ and $f_{\rm EDE}$, one can fix $M$.}
\begin{align}
	\{ \e, ~\lambda, ~m, ~g, ~\phi_0, ~\frac{\alpha}{f_a}, ~f_{\rm EDE},~ z_c\} \,.
\end{align}
We acknowledge that our model's inclusion of 6 additional parameters, compared to previous EDE models, presents a downside in certain aspects. However, the dynamic and natural setup remains an appealing aspect of our idea. While simplicity, often embodied in models with fewer parameters, tends to offer elegance and ease of interpretation, the complexity inherent in models with more parameters can often capture the subtleties and complexities of a phenomenon more comprehensively. The nonlinear dynamics of symmetry breaking within our model render it notably rich, demanding further exploration. 
%and extensive brainstorming to fully grasp its core concepts and potential implications.

%We have identified the viable model parameters where the Hubble tension is solved by EDE and the dark photons explain a non-negligible portion of the observed DM for a wide range of the parameter space. We also found that the parameter space where the dark photons make total observed DM. While our semi-analytical analysis indicates the model's potential to address both issues, it appears that they cannot be simultaneously resolved in our setup. A comprehensive lattice simulation and a series of MCMC runs are imperative for a more detailed understanding. We defer these simulations to future studies, aiming to explore the broader parameter space encompassing the symmetry-breaking mechanism.

The proposed model unveils numerous intriguing avenues for the early Universe. Among the foremost inquiries are the origins of the waterfall and axion fields. In the context of the standard model, one can assume that these fields were emerged during the reheating process alongside particles. The fate of the axion field is also important, potentially influencing late-time dynamics. Furthermore, cosmic strings may exert an impact on subsequent structure formation. Additionally, the setup harbors the potential to generate gravitational waves falling within the sensitivity ranges of the current and upcoming GW observatories. We intend to explore these aspects further in future works.

\vspace{1cm}

{\bf Acknowledgments:} 
%We would like to thank ....
The author is grateful to Hassan Firouzjahi, Shinji Mukohyama, and Borna Salehian for their valuable remarks and for fruitful discussions. We acknowledge Ali Akbar Abolhasani and Mohammad Hossein Namjoo for insightful discussions and comments. 
%A.T. is grateful to the Yukawa Institute for Theoretical Physics (YITP) for their hospitality during early stage of this work. 
%The authors 
%%%%%%%%%%%%%%%%%%%%%%%%%%%%
%A. T. thanks the Yukawa Institute for Theoretical Physics at Kyoto University, where this work was initiated %[completed] 
%during the stay under "Visitors Program of FY2022".
%%%%%%%%%%%%%%%%%%%%%%%%%%%%
%The authors thank 
A. T. thanks the Yukawa Institute for Theoretical Physics at Kyoto University. Discussions during the stay under "Visitors Program of FY2022" were useful to complete this work.
We would like to thank University of Rwanda, EAIFR, and ICTP for their kind hospitalities when some parts of the project were in hand.
%Also, we thank ... for his helpful comments. 
%Mohammad Hossein Namjoo for helpful discussions and comments. 
%during the 17th international workshop on the ”Dark Side of the Universe” 
%SM is grateful for the hospitality of Perimeter Institute, the cosmology group at Simon Fraser University and the Theoretical Physics Institute at University of Alberta, where part of this work was carried out.

\appendix

\iffalse
\section{How fast is the transition?}
\textcolor{blue}{Assuming that EDE furnish a fraction $f_{\rm EDE}$ of the total energy at that time, the condition $g^2 \psi_{\rm min}^2 \gg H_{\rm c}^2$, requires $\alpha g^2 \gg (\frac{M}{M_P})^2$. }
\fi

\section{Upper bound on the parameter ${\cal C}$}
\label{C_upper_bound}
At the end of EDE, we have
\begin{align}
%\rho_{\rm c}^{(r)}=
\rho^{(r)}(a_{\rm c}) &= \rho^{(r)}_{\rm eq}\left(
\dfrac{a_{\rm c}}{a_{\rm eq}}
\right)^{-4} \,,
\\
%\rho_{\rm c}^{(m)}=
\rho^{(m)}(a_{\rm c}) &= \rho^{(m)}_{\rm eq}\left(
\dfrac{a_{\rm c}}{a_{\rm eq}}
\right)^{-3} \,.
\end{align}
Considering $\rho^{(m)}_{\rm eq}=\rho^{(r)}_{\rm eq}$, we find
\begin{align}
\rho^{(m)}(a_{\rm c})=\rho^{(r)}(a_{\rm c}) ~ \dfrac{a_{\rm c}}{a_{\rm eq}} \,.
\end{align}
Now we consider that after the EDE phase the total energy of Universe comes from radiation and matter,
\begin{align}
\rho(a_{\rm c}) = %\rho^{(m)}(a_{\rm c})+\rho^{(r)}(a_{\rm c}) =\rho^{(r)}(a_{\rm c}) \left(1+\dfrac{a_{\rm c}}{a_{\rm eq}}\right)= 
\dfrac{\pi^2}{30}g_{*{\rm c}} T_{\rm c}^4\left(1+\dfrac{a_{\rm c}}{a_{\rm eq}}\right) \,.
\end{align}
We also assume the matter component includes two parts: %$\rho^{(m)}(a_{\rm c})= \rho^{(A)}(a_{\rm c})+\rho^{(B)}(a_{\rm c})$; 
non-dark photon part $\rho^{(B)}(a_{\rm c})={\cal D} \rho(a_{\rm c})$ and dark photon $\rho^{(A)}(a_{\rm c})={\cal C} \rho(a_{\rm c})$. Hence we arrive at
\begin{align}
\label{C+D}
{\cal D}+{\cal C} = \dfrac{a_{\rm c}}{a_{\rm c}+a_{\rm eq}} \,. %\simeq\dfrac{z_{\rm eq}}{z_{\rm eq}+z_{\rm c}} \simeq 0.4
\end{align}
As an example, for $z_{\rm eq} \simeq 3400$ and $z_{\rm c}\simeq 5000$, it implies that at $z_{\rm c}$, the matter components make up approximately $40\%$ of the total energy density, allowing for a model with minimal deviation from $\Lambda$CDM cosmology. 
%A relevant question here is that: what is the minimum value of ${\cal D}$ to have a consistent BBN scenario?

%%%%%%%%%%%%%%%%%%%%%%%%%%%%%%%%%%%%%%%%%%%%%%%%%%%%%%%%%%%%%%%%%%%%%%%%%%%%%%%%%%%%%%%%%%%%%%%%%%%%%%%%%%%%%%%%%%%%%%%%%%%%%%%%%%%%%%%%%%%%%%%%

\vspace{0.7cm}

%%%%%%%%%%%%%%%%%%%%%%%%%%%%%%%%%%%%%%%%%%%%%%%%%%%%%
%\begin{thebibliography}{}
%
%\end{thebibliography}{}
\small
%\bibliography{WhiteHole}{}
%\bibliography{references}
%\bibliographystyle{JHEP}
%\bibliographystyle{JHEPNoTitle}
		
\end{document}